\documentclass[prc,twocolumn,nofootinbib]{revtex4-1}

\usepackage{graphicx}
\usepackage{dcolumn}
\usepackage{bm}
\usepackage{slashed}
\usepackage{amsmath,graphicx}
\usepackage[colorlinks=true,linktocpage=true,linkcolor=blue,citecolor=blue]{hyperref}
\usepackage{float}
\usepackage{nicefrac}
\usepackage[normalem]{ulem}
\usepackage{amsmath}
\usepackage{subfigure}

\begin{document}

\def\AH{\small \hbox{AHYDRO }}
\def\MIS{\small \hbox{BRSSS }}
\def\DNMR{\small \hbox{DNMR }}
\def\IS{\small \hbox{MIS }}
\def\RTA{\small \hbox{RTA }}
\def\QCD{\small \hbox{QCD }}
\def\QGP{\small \hbox{QGP }}

\def\chieps{\chi_\varepsilon\!\left(\frac{\mu}{T}\right)}
\def\chiepsinit{\chi_\varepsilon\!\left(\frac{\mu_0}{T_0}\right)}
\def\chiepsp{\chi_\varepsilon\!\left(\frac{\mu^\prime}{T^\prime}\right)}

\def\chiden{\chi_n\!\left(\frac{\mu}{T}\right)}
\def\chideninit{\chi_n\!\left(\frac{\mu_0}{T_0}\right)}
\def\chidenp{\chi_n\!\left(\frac{\mu^\prime}{T^\prime}\right)}

\def\tel{\tau_{\rm el}}
\def\telprime{\tau_{\rm el}^\prime}
\def\telbis{\tau_{\rm el}^{\prime \prime}}
\def\tf{{\tau_{\infty}}}
\def\fst{f_{\rm st}(\mu,T)}
\def\finf{f_\infty}

\def\tinprime{\tau_{\rm in}^\prime}
\def\tinbis{\tau_{\rm in}^{\prime \prime}}

\def\mup{\mu^{\prime}}
\def\Tp{{T^{\prime}}}

\def\vpper{{\bf p}_\perp}
\def\ppar{p_\parallel}

\def\eps{\varepsilon}
\def\Ppar{P_\parallel}
\def\Pper{P_\perp}
\def\Peq{P_{\rm eq}}


\def\teq{\tau_{\rm eq}}
\def\t0{{\tau_0}}

\def\calR{{\cal R}}
\def\calRL{{\cal R}_\parallel}

\def\feq{f_{\rm eq}} 
\def\fa{f_{\rm aniso}} 

\def\tn{t_{n}}
\def\xin{\xi_{n}}

\def\pper{p_\perp}

\def\wpar{w_\parallel}

\newcommand{\rf}[1]{Eq.~(\ref{#1})}
\newcommand{\rfn}[1]{~(\ref{#1})}

\newcommand{\rfs}[1]{Sec.~\ref{#1}}
\newcommand{\rff}[1]{Fig.~\ref{#1}}
\newcommand{\rfc}[1]{Ref.~\cite{#1}}

\newcommand{\beq}{\begin{eqnarray}}
\newcommand{\eeq}{\end{eqnarray}}

\newcommand{\bea}{\begin{eqnarray}}
\newcommand{\beal}[1]{\begin{eqnarray}\label{#1}}
\newcommand{\eea}{\end{eqnarray}} 

\newcommand{\be}{\begin{equation}} 
\newcommand{\bel}[1]{\begin{equation}\label{#1}}
\newcommand{\ee}{\end{equation}} 

\newcommand{\p}{\partial}
\newcommand{\f}[2]{\frac{#1}{#2}}
\newcommand{\nn}{\nonumber}

\newcommand{\ms}[1]{{\textcolor{blue}{#1}}}

 
\title{Gradient expansion for anisotropic hydrodynamics}

\author{Wojciech Florkowski}

\affiliation{Jan Kochanowski University,  PL-25-406 Kielce, Poland}

\author{Radoslaw Ryblewski}

\affiliation{Institute of Nuclear Physics, Polish Academy of Sciences, PL-31-342 Krak\'ow, Poland}

\author{Micha\l{} Spali\'nski}
\affiliation{University of Bia\l{}ystok, PL-15-245
  Bia\l{}ystok, Poland and National Center for Nuclear Research, PL-00-681
  Warsaw, Poland} 

\date{\today}

\begin{abstract}
We compute the gradient expansion for anisotropic hydrodynamics. The results are compared with the corresponding expansion of  
the underlying kinetic-theory model with the collision term treated in the relaxation time approximation.
We find that a recent formulation of anisotropic hydrodynamics based on an anisotropic matching principle
yields the first three terms of the gradient expansion in agreement with those obtained for the kinetic theory. This gives further support
for this particular hydrodynamic model as a good approximation of the
kinetic-theory approach. We further find that the gradient expansion of anisotropic hydrodynamics is 
an asymptotic series, and  the singularities of the analytic continuation of its Borel
transform indicate the presence of non-hydrodynamic modes.
\end{abstract}



\maketitle 

\section{Introduction}
\label{sect:intro}

Theories of relativistic hydrodynamics have enjoyed a remarkable decade of
advances both in the theoretical aspects of their formulation and in
their numerical implementation. The driving force behind these developments
was the successful application of this type of description to the physics of the quark-gluon
plasma (QGP). During this time a number of different theories of relativistic
hydrodynamics have been formulated, and one of the aims of this paper is to
clarify how they are related.  We adopt the point of view that computing
gradient expansions of the energy-momentum tensor provides a natural means of
making such comparisons. It is also a direct way to compare the effective,
hydrodynamic description with an underlying microscopic theory or model.

Our main focus of interest here are theories of anisotropic 
hydrodynamics (\AH\!\!)~\cite{Florkowski:2010cf,Martinez:2010sc}, whose form
looks quite different from other approaches. One of our goals is to use the
gradient expansion as a way to clarify similarities and differences 
which may not be obvious at first glance. Another major motivation of this
work is to compare the leading terms in the gradient expansion of the
energy-momentum tensor of \AH  with the
corresponding terms calculated recently directly at the level of kinetic theory
\cite{hks}.   To this end,  we calculate the gradient expansion for anisotropic 
hydrodynamics in the special case of boost-invariant and
transversely homogeneous systems that have been the focus of many studies in
recent years in the context  of early thermalisation and hydrodynamization of matter produced in heavy-ion
collisions~\cite{Chesler:2009cy,Heller:2011ju,Heller:2012km,Jankowski:2014lna,Gelis:2013rba,Berges:2013eia,Ryblewski:2013eja,Ruggieri:2015yea}.

We analyse two different formulations of anisotropic
hydrodynamics~\cite{Tinti:2013vba,Tinti:2015xwa}, as well as 
M\"uller-Israel-Stewart viscous hydrodynamics
(\IS\!\!)~\cite{Muller:1967zza,Israel:1976tn,Israel:1979wp}, and especially
its modern version~\cite{Baier:2006um}, which we will refer to as \MIS\!\!. We compare the
corresponding gradient expansions with the expansion obtained for the
kinetic-theory model with the collision term treated in the relaxation time
approximation (\RTA\!\!)~\cite{hks}.

A comparison of the gradient expansions provides an indication of how successful the
various hydrodynamic approaches are in reproducing close-to-equilibrium dynamics
governed by the underlying microscopic theory. We find that the formulation of
anisotropic hydrodynamics based on the {\it anisotropic matching
  principle}~\cite{Tinti:2015xwa} yields the first three terms in the gradient
expansion which agree exactly with the terms obtained for the kinetic
theory. This gives support for this particular anisotropic hydrodynamics
description as a good approximation   for the underlying kinetic-theory model.

From a wider perspective, this suggests that effective hydrodynamic descriptions
tailored to a specific microscopic theory may provide a better picture for a
given system than a general framework such as the \MIS theory. The latter
provides a universal set of equations valid for all relativistic systems
sufficiently close to equilibrium.  Different microscopic theories, such as
Quantum Chromodynamics (\QCD\!\!), ${\cal N }$=\,4 supersymmetric Yang-Mills
theory, or kinetic-theory models, are distinguished by different sets of
transport coefficients. However, the gradient expansions are guaranteed to match only up to second
order, beyond which they differ. In contrast, anisotropic hydrodynamics, as understood
here, aims to provide a structure more closely attuned to the kinetic theory in the
RTA, which is the reason why it can provide a better description.

We also examine the large order behaviour of the gradient series generated by
the anisotropic hydrodynamic theories and find that they are asymptotic, with
the coefficients $g_n$ growing as $n!$. The Borel transform technique applied
to this series indicates the presence of purely damped non-hydrodynamic
modes. This parallels earlier findings in the \MIS theory~\cite{Heller:2015dha}.

The paper is organised as follows: In Sec.~\ref{sect:hydro} we discuss the basic
theoretical structures of perfect-fluid and viscous relativistic
hydrodynamics. We distinguish between the notion of hydrodynamic expansion
that is used to construct hydrodynamic equations from the underlying kinetic
theory~\cite{Denicol:2012cn,Denicol:2012es,Denicol:2014loa} 
and the gradient expansion which is a formal, infinite, expansion of the
energy-momentum tensor of a given theory around the perfect fluid form.
In Sec.~\ref{sect:aniso} we introduce a reorganised hydrodynamic expansion
that leads to the concept of anisotropic hydrodynamics. The constraints resulting
from imposing longitudinal boost invariance and transverse homogeneity are
implemented in Sec.~\ref{sect:binv}.
The kinetic theory model considered in this paper (based on the
relaxation time approximation) is shortly described in Sec.~\ref{sect:rta}.
 In Sec.~\ref{sect:aniso2} we present two different versions of equations of the
anisotropic hydrodynamics. In the first case one uses moments of the Boltzmann
equation~\cite{Tinti:2013vba}, while in the second case one uses the
anisotropic matching principle combined with the exact treatment of the
dynamical equations for the pressure corrections~\cite{Tinti:2015xwa}. The latter
method introduces an infinite set of the coupled equations that is,
eventually, truncated at the leading order. The M\"uller-Israel-Stewart
approach to hydrodynamics is discussed in Sec.~\ref{sect:mis}. Our main
results on the gradient expansion for anisotropic hydrodynamics are presented in
Secs.~\ref{sect:gradaniso} and \ref{sect:Borel}. We summarize and conclude in
Sec.~\ref{sect:summary}.

Throughout the paper we use natural units with $c=k_B=\hbar=1$. The metric
tensor is $g_{\mu\nu}= \hbox{diag}(1,-1,-1,-1).$

\section{Theories of Relativistic Hydrodynamics}
\label{sect:hydro}

\subsection{Dynamical variables and evolution equations}
\label{sect:hydrovars}

By a hydrodynamic description one means a theoretical framework that uses a
small set of {\it fluid variables}. For a {\it perfect fluid} these can be
chosen as the local energy density $\eps(x)$ and local hydrodynamic flow
vector $U^\mu(x)$, which is normalized as $U^2=1$. To set the stage for the
developments described in the remainder of the paper, it is worthwhile to
briefly review the basic conceptual structures used to formulate theories of
relativistic hydrodynamics. 

The point of departure is the assumption that we are dealing with a system
which reaches global thermodynamic equilibrium at late times. At this stage one can
be agnostic about the fundamental physics governing this system: it could be
composed of well defined quasiparticles, but it need not be.

The equilibrium energy-momentum tensor in the rest-frame is given by
\bel{equilibrium}
T^{\mu\nu}_{\rm EQ} = \hbox{diag}\left(\eps_{\rm EQ}, P(\eps_{\rm
  EQ}), P(\eps_{\rm EQ}), P(\eps_{\rm EQ})\right),
\ee
where we assume that the equation of state is known, so that the pressure $P$
is a given function of the energy density $\eps_{\rm EQ}$. It is worth
stressing, that $T^{\mu\nu}_{\rm EQ}$ is a classical object which we should identify with the
expectation value of the energy-momentum tensor operator in the underlying
quantum theory.

The components of the equilibrium energy-momentum tensor\rfn{equilibrium} can
be written in an arbitrary boosted frame of reference as
\begin{eqnarray}
T^{\mu\nu}_{\rm EQ} = \varepsilon_{\rm EQ} U^\mu U^\nu - P(\varepsilon_{\rm EQ}) \Delta^{\mu
  \nu}, 
\label{TmunuEQ}
\end{eqnarray}
where $U^\mu$ is a constant boost velocity, and $\Delta^{\mu\nu}$ is the
operator that projects on the space orthogonal to $U^\mu$, namely
\begin{equation}
\Delta^{\mu\nu} = g^{\mu\nu}-U^\mu U^\nu.
\label{defDelta}
\end{equation}
Of course, the four-vector $U^\mu$ can equally well be regarded as a constant fluid
four-velocity.

The energy-momentum tensor of a perfect fluid is obtained by allowing
the variables $\eps$ and $U^\mu$ to depend on the spacetime point $x$.
In this way one obtains
\begin{eqnarray}
T^{\mu\nu}_{\rm eq} = \varepsilon(x) U^\mu(x) U^\nu(x) - P(\varepsilon(x))
\Delta^{\mu \nu}(x).
\label{TmunuPerf}
\end{eqnarray}
In this equation, and in those which follow, the subscript ``eq'' refers
to local thermal equilibrium. 

It is often convenient to introduce local effective temperature $T(x)$ by the
condition that the equilibrium energy density at this temperature agrees with
the non-equilibrium value of the energy density, namely
\begin{eqnarray}
\eps_{\rm EQ}(T(x))  = \eps_{\rm eq}(x) = \eps(x).
\label{LM0}
\end{eqnarray}
One can then express the perfect fluid energy-momentum tensor in terms of the
fluid variables $T(x)$ and $U^\mu(x)$.  Note that the relativistic perfect fluid
energy-momentum tensor\rfn{TmunuPerf} is the most general symmetric tensor
which can be expressed in terms of these variables without using derivatives.

The dynamics of the perfect fluid theory is provided by the conservation
equations of the energy-momentum tensor
\bel{conservation}
\p_\mu T_{\rm eq}^{\mu\nu} = 0 .
\ee
These are four equations for the four independent hydrodynamic fields, that
form a self-consistent hydrodynamic theory. 

The essential physical element which is missing in the approach based
on~Eqs.~(\ref{TmunuPerf}) and (\ref{conservation}) is dissipation. To account
for it, we have to introduce correction terms to $T_{\rm eq}^{\mu\nu} $ and
write the complete energy-momentum tensor components $T^{\mu\nu}$ as
\begin{eqnarray}
T^{\mu\nu} = T_{\rm eq}^{\mu\nu} + \Pi^{\mu\nu}.
\label{Tmunu}
\end{eqnarray}
Here (see, e.g., \cite{Bhattacharya:2011eea}) one can impose the condition
$\Pi^{\mu\nu} U_\nu=0$, which corresponds to the Landau definition of the
hydrodynamic flow $U^\mu$ by the formula
\begin{eqnarray}
T^\mu_{\,\,\,\nu} U^\nu = \varepsilon\, U^\mu.
\label{LandFrame}
\end{eqnarray}
It proves useful to further decompose $\Pi^{\mu\nu}$ into two components,
\bel{pidecomp}
\Pi^{\mu\nu} = \pi^{\mu\nu} - \Pi \Delta^{\mu \nu},
\ee
which introduces the bulk viscous pressure $\Pi$ (the trace part of
$\Pi^{\mu\nu}$) and the shear stress tensor $\pi^{\mu\nu}$ which is symmetric,
$\pi^{\mu\nu}=\pi^{\nu\mu}$, traceless, $\pi^{\mu}_{\,\,\,\mu}=0$, and
orthogonal to $U^\mu$, $\pi^{\mu\nu} U_\nu=0$.

Equation (\ref{Tmunu}) encodes ten independent components of $T^{\mu\nu}$ in
terms of the effective temperature, three independent components of $U^\mu$,
five independent components of  $\pi^{\mu\nu}$, and the bulk viscous pressure 
$\Pi$. We note that the latter vanishes for conformal systems, for which the
entire energy-momentum tensor is traceless. 

We still have the four conservation equations at our disposal, but to obtain a closed system of 
equations one needs additional information. 
The most straightforward option is to express $\Pi^{\mu\nu}$ in terms of the hydrodynamic
variables and their gradients. Since the perfect-fluid energy-momentum tensor
contains no gradients, it is natural to try to build up the theory as a series of
corrections in gradients. The simplest possibility is to include terms with only a single gradient,
which leads to the relativistic Navier-Stokes theory~\cite{LandauHyd}, in which the bulk pressure and
shear stress tensor are given by the gradients of the flow vector 
\begin{eqnarray}
\Pi = -\zeta\, \partial_\mu U^\mu, \quad  \pi^{\mu\nu} = 2 \eta  \sigma^{\mu\nu}.
\label{NavierStokes}
\end{eqnarray}
Here $\zeta$ and $\eta$ are the bulk and shear viscosity coefficients,
respectively, and $\sigma^{\mu\nu}$ is the shear flow tensor defined as 
\begin{eqnarray}
\sigma^{\mu\nu} = \Delta^{\mu\nu}_{\alpha \beta} \partial^\alpha U^\beta,
\end{eqnarray}
where the projection operator $ \Delta^{\mu\nu}_{\alpha \beta} $ has the form
\begin{eqnarray}
 \Delta^{\mu\nu}_{\alpha \beta}  = \frac{1}{2} \left(
 \Delta^\mu_{\,\,\,\alpha}  \Delta^\nu_{\,\,\,\beta} +  \Delta^\mu_{\,\,\,\beta}  \Delta^\nu_{\,\,\,\alpha}  \right) 
 - \frac{1}{3}  \Delta^{\mu\nu} \Delta_{\alpha \beta}.
\end{eqnarray}
While the evolution equations obtained in this way are covariant, they have
solutions which propagate with arbitrarily high velocities, leading to
causality violation and instabilities in numerical
simulations~\cite{kostadt,Romatschke:2009im}. 

Adding extra terms with higher gradients on the right-hand side of
(\ref{NavierStokes}) does not help to solve the problem with causality.  The
only known way to avoid it is to relax the assumption that $\Pi$ and
$\pi^{\mu\nu}$ are expressed locally in terms of the hydrodynamic variables
$T, U^\mu$, and (a finite number of) their spacetime
derivatives~\cite{Muller:1967zza}. This means however, that the conservation
equations alone are no longer enough to determine the dynamics of $T^{\mu\nu}$
and one needs to postulate additional dynamic equations, or derive them
(possibly by some heuristic means). The outcome, a closed set of hydrodynamic
equations, will clearly involve additional degrees of freedom beyond those
already present in the theory of the perfect fluid. To write down such
equations will require additional assumptions, or additional information
beyond what is embodied by conservation laws.

\subsection{Approaches to finding evolution equations}
\label{sect:hydro_exp}

A well-known and widely applied approach to the task of positing a set of
closed equations for the hydrodynamic fields is the M{\"u}ller-Israel-Stewart
theory~\cite{Muller:1967zza,Israel:1976tn,Israel:1979wp}, in its modern 
incarnation described in \rfc{Baier:2006um} (\MIS\!\!). This approach (discussed further
in \rfs{sect:mis}) basically parameterizes dominant contributions classified
by symmetries and the number of gradients. It does not make any special
assumptions about the microscopic dynamics, which accounts for its
generality. A major advantage of this approach is that the resulting equations
are causal at least for some domain in the space of transport coefficients.

If we commit to a specific microscopic model, we gain the option of deriving
(at least in a heuristic way) a set of hydrodynamic equations which can
provide a better physical picture than a generic approach such as the \MIS
theory. An important testbed for this idea is provided by kinetic theory with
an idealized collision kernel, further discussed below in
Sec.~\ref{sect:rta}.  In this case a number of complementary approaches exist,
which we now briefly review.

One important approach makes use of the {\it hydrodynamic expansion}: this is
the process of constructing the dynamical equations order by order in the
Knudsen and inverse Reynolds
numbers~\cite{Denicol:2012cn,Denicol:2012es,Denicol:2014loa}.  The
hydrodynamic expansion is performed around the local equilibrium state that
corresponds to the perfect-fluid limit~$T^{\mu\nu}_{\rm eq}$. The Knudsen
number is the ratio of the molecular mean free path length to a representative
physical length scale. On the other hand, the inverse Reynolds number
describes deviations of the energy-momentum components from their local
equilibrium values --- they are typically expressed by the ratios
$\sqrt{\pi^{\mu\nu} \pi_{\mu\nu}}/P$ and $\Pi/P$.  The hydrodynamic expansion 
serves as a tool to systematically derive hydrodynamic equations from kinetic theory.

Another approach to the task of formulating a closed set of hydrodynamic
equations for models of kinetic theory is known under the name of anisotropic
hydrodynamics~\cite{Florkowski:2010cf,Martinez:2010sc} (for a recent review
see \cite{Strickland:2014pga}).  This name originated in the desire of finding
hydrodynamic equations suited to describing early stages of evolution of the
quark-gluon plasma (\QGP\!\!) produced in heavy-ion collisions. Equations of
anisotropic hydrodynamics were formulated in such a way as to capture some
features of highly anisotropic initial states, but also to ensure that at late
times their predictions should be consistent with \MIS\!\!. In modern
formulations, the equations of anisotropic hydrodynamics are suitable for
studying arbitrary flows.  Further details of this approach (which is central
to the present article) are given in \rfs{sect:aniso}.

\subsection{Gradient expansions}
\label{sect:grad_exp}

The gradient expansion is an effective way to quantify the approach to
equilibrium as well as to gather information about the non-hydrodynamic
sector. This latter information is attainable by considering perturbations
around the hydrodynamic solution, or through the study of large order behavior
of the gradient series, with the former being encoded in the leading terms.

As briefly reviewed in \rfs{sect:hydrovars}, once dissipative effects are
incorporated within a hydrodynamic framework, one loses the universality of
perfect fluid theory, and many different sets of hydrodynamic equations are
possible. This raises the question of how they are to be compared, and how can
they be reconciled with computations carried out directly in the microscopic
theory.

One measuring stick that can be used to survey this realm of
possibilities is the gradient expansion: a formal, infinite, expansion of the
energy-momentum tensor $T^{\mu\nu}$ around the perfect fluid
form~$T^{\mu\nu}_{\rm eq}$ in powers of gradients of the fluid variables
$T(x)$ and $U^\mu(x)$:
\begin{eqnarray}
\hspace{-0.5cm} T^{\mu\nu} &=& T^{\mu\nu}_{\rm eq} + \hbox{powers of gradients of } T \hbox{ and } U^\mu . 
\label{TmunuGrad}
\end{eqnarray}
Such an expression can arise in calculations based on microscopic models (such
as the AdS/CFT representation of ${\cal N}$=\,4 supersymmetric Yang-Mills
theory \cite{Bhattacharyya:2008jc} or kinetic theory \cite{hks}).  
Crucially, it also arises from any set of hydrodynamic equations as a generic
late-time solution. Given the set of evolution equations for the shear stress
tensor, we can always find such a solution by writing down the most general
gradient series consistent with Lorentz symmetry (and any other constraints,
such as perhaps conformal invariance), and determine the scalar coefficient
functions by using the evolution equation order by order in gradients.

We wish to stress the need to make a clear distinction between the gradient
expansion discussed here and the hydrodynamic expansions discussed in the
previous subsection as a means of deriving {\em hydrodynamic  equations} from
kinetic theory.  As discussed there, various methods exist for writing down
hydrodynamic equations; once some set of hydrodynamic equations is found, 
one can look for a formal solution in the form of a gradient expansion of the
energy-momentum tensor as in \rf{TmunuGrad}. 

By comparing such formal solutions, one may quantify differences between
different hydrodynamic theories, as well as compare them to a given
microscopic model. This process also determines any phemomenological
parameters appearing at the level of hydrodynamics in terms of parameters
appearing in the fundamental theory.

\section{Anisotropic hydrodynamics -- reorganized hydrodynamic expansion}
\label{sect:aniso}


As discussed above, in the standard approach to viscous hydrodynamics, the
energy-momentum tensor components $T^{\mu\nu}$ may be treated as functions of
$T$, $U^\mu$, $\pi^{\mu\nu}$, and $\Pi$, which we write schematically as
\beq
T^{\mu\nu} &=&  T^{\mu\nu}\left(T,U^\mu,\pi^{\mu\nu},\Pi\right) \nonumber \\
&=& T^{\mu\nu}_{\rm eq} (T,U) + \pi^{\mu\nu} - \Pi  \Delta^{\mu \nu} \nonumber \\
&\equiv& T^{\mu\nu}_{\rm eq} + \delta T^{\mu\nu}.
\label{TmunuDecEq}
\eeq
The hydrodynamic equations that determine the dynamics of~$T^{\mu\nu}$ contain
various terms that may be characterised by the power of space-time gradients
and/or the power of the dissipative terms they include~\footnote{Strictly
  speaking one considers the powers of the ratios $\pi^{\mu\nu}/P$ and $\Pi/P$
  that are known as the inverse Reynolds numbers.}. For example, in the first
order of viscous hydrodynamics one deals with $\pi^{\mu\nu}$ and $\Pi$ and
also with the gradients of $T$ and $U^\mu$, see Eqs.~(\ref{NavierStokes}). In
the second order, the products of $\pi^{\mu\nu}$ and $\Pi$ appear, as well as
the gradients of $\pi^{\mu\nu}$ and $\Pi$. Such approach may be continued to
higher orders but, in practical applications, one stops at the third order
(for example, see \cite{Jaiswal:2014raa}).

Anisotropic hydrodynamics can be treated as a method to reorganize this kind of expansion within the framework of the kinetic theory. Thus, from now on we use the concepts of the phase space distribution function $f(x,p)$ and express different physical quantities as the moments of $f(x,p)$ (in the three-momentum space). Within \AH one separates the description of viscous effects into two parts.  The first part is characterised by the new fluid variables $\xi^{\mu\nu}$ and $\phi$, see 
Refs.~\cite{Martinez:2012tu,Tinti:2013vba,Nopoush:2014pfa,Tinti:2014yya,Tinti:2015xra,Tinti:2015xwa}. They may account for large possible values of the shear stress tensor and bulk viscosity and should be treated in a non-perturbative manner, similarly to $T$ and $U^\mu$. The second part is characterised by the tensor ${\tilde \pi}^{\mu\nu}$ and ${\tilde \Pi}$ that are treated similarly as $\pi^{\mu\nu}$ and $\Pi$ in the standard case~\cite{Bazow:2013ifa,Bazow:2015cha,Molnar:2016vvu,Molnar:2016gwq}. Thus, we write
\beq
T^{\mu\nu} &=&  T^{\mu\nu}\left(T,U^\mu,\xi^{\mu\nu},\phi,{\tilde \pi}^{\mu\nu},{\tilde \Pi}\right) \nonumber \\
&=& T^{\mu\nu}_{\rm a} (T,U,\xi^{\mu\nu},\phi) + {\tilde \pi}^{\mu\nu} - {\tilde \Pi}  \Delta^{\mu \nu} \nonumber \\
&\equiv& T^{\mu\nu}_{\rm a} + \delta {\tilde T}^{\mu\nu}.
\label{TmunuDecA}
\eeq

Several comments are in order now:
\begin{itemize}

\item[i)] Introducing the fluid variables $\xi^{\mu\nu}$ and $\phi$ together with ${\tilde \pi}^{\mu\nu}$ and ${\tilde \Pi}$ means that the ``viscous'' degrees of freedom may be doubled and we need more than typical ten hydrodynamic equations to determine dynamics of the energy-momentum tensor. This can be easily achieved within kinetic theory approach by including, for example, a sufficient number of the moments of the kinetic equation. The selection of the moments is, however, not well defined. One considers usually the lowest possible moments~\cite{Bazow:2013ifa}, since they are most sensitive to the low momentum sector which is expected to be well described by hydrodynamics-like models. We come back to the discussion of this ambiguity below in the point vii).

\item[ii)] The tensor  $\xi^{\mu\nu}$ has similar geometric properties as  $\pi^{\mu\nu}$, namely, it is symmetric, transverse to $U^\mu$ and traceless~\cite{Tinti:2014yya,Tinti:2015xra,Tinti:2015xwa}. This means that it has in general five independent components. However, in practical applications one often uses a simplified versions of $\xi^{\mu\nu}$ that contains one or two independent parameters. In such cases only these degrees of freedom may be ``doubled''. We note that the use of simplified forms of  $\xi^{\mu\nu}$ is very often a consequence of the system's symmetries such as boost invariance, homogeneity in the transverse plane or cylindrical symmetry.

\item[iii)] The tensor   ${\tilde \pi}^{\mu\nu}$ is also symmetric, transverse to $U^\mu$ and traceless. Consequently, using the Landau frame, one can determine the effective temperature of the system $T$ and the flow $U^\mu$ by the equation
\begin{eqnarray}
\varepsilon_{\rm EQ} (T(x))\, U^\mu(x) = T^{\mu \nu}_{\rm{a}}(x) U_\nu(x).
\label{LandFrameA}
\end{eqnarray}

\item[iv)] Since the substantial part of the viscous effects is included with
  the help of the variables $\xi^{\mu\nu}$ and $\phi$, one expects that the
  terms ${\tilde \pi}^{\mu\nu}$ and ${\tilde \Pi}$ are small compared to the
  equilibrium pressure $P$. The expansion in the ratios
   $\sqrt{{\tilde  \pi}^{\mu\nu} {\tilde  \pi}_{\mu\nu}}/P$ and ${\tilde \Pi}/P$ is discussed in this context as an
  expansion in the modified inverse Reynolds
  numbers~\cite{Bazow:2013ifa,Bazow:2015cha,Molnar:2016vvu,Molnar:2016gwq}.

\item[v)] Using the kinetic theory approach, Eq.~(\ref{TmunuDecA}) is reproduced with the distribution function that has a structure
\begin{eqnarray}
f(x,p) = f_{\rm a}(x,p)  + \delta {\tilde f}(x,p).
\label{deltafa}
\end{eqnarray}
Here $f_{\rm a}(x,p)$ is the anisotropic distribution function in the momentum
space. It can be regarded as an extension of the equilibrium distribution
$f_{\rm eq}(x,p)$, which depends not only on $T$ and $U^\mu$ but also on
$\xi^{\mu\nu}$ and $\phi$. In the limit $\xi^{\mu\nu}, \phi \to 0$ one finds
that $f_{\rm a}(x,p) \to f_{\rm eq}(x,p)$. Two special forms of $f_{\rm
  a}(x,p)$ will be discussed below, see Eqs.~(\ref{RS1})~and~(\ref{TF1}).

\item[vi)]  In the leading order of anisotropic hydrodynamics, we neglect the corrections 
$ \delta {\tilde T}^{\mu\nu}$ in (\ref{TmunuDecA}) and $\delta {\tilde f}(x,p)$ in (\ref{deltafa}). 
The complete energy-momentum tensor has the form
\begin{eqnarray}
T^{\mu \nu} = T_{\rm a}^{\mu \nu} = k \int \frac{d^3p}{(2\pi)^3 \,p^0} \,\,p^\mu p^\nu f_{\rm a}(x,p),
\label{TmunuA}
\end{eqnarray}
where $k$ is a degeneracy factor. In this case, the components of $T_{\rm a}^{\mu \nu}$ depend, 
in general, on ten independent parameters contained in the set: $T$, $U^\mu$, $\xi^{\mu\nu}$ and
$\phi$. The equations of anisotropic hydrodynamics specify the dynamics of $T_{\rm
  a}^{\mu \nu}$. They include four equations that follow directly from the
energy-momentum conservation law and additional six equations that should be
derived from some other (microscopic) theory.

\item[vii)] Doubling of the viscous degrees of freedom can be avoided if the
  use of a certain parameter in the set ($\xi^{\mu\nu}, \phi$) is accompanied
  with the elimination of some parameter in the set (${\tilde \pi}^{\mu\nu},
  {\tilde \Pi}$). For example, using the bulk parameter $\phi$ we can set $
  {\tilde \Pi}=0$. The extreme strategy in this context is to assume that the
  parameters $\xi^{\mu\nu}$ and $\phi$ are chosen in such a way that
\begin{eqnarray}
T^{\mu \nu} = T_{\rm a}^{\mu \nu}.
\label{TmunuAP}
\end{eqnarray}
This formula represents the {\it anisotropic matching principle} introduced by Tinti~\cite{Tinti:2015xwa}. We note that (\ref{TmunuAP}) is formaly equivalent to (\ref{TmunuA}), however, it is obtained with different assumptions: instead of neglecting the term $ \delta {\tilde f}(x,p)$ in (\ref{deltafa}) one assumes that $ \delta {\tilde f}(x,p)$ might be finite but it does not contribute to $T^{\mu \nu}$. 

\end{itemize}

We can now define the gradient expansion for the leading-order anisotropic
hydrodynamics. Given $T(x)$ and $U^\mu(x)$ we define $f_{\rm eq}(x,p) $ and
$T^{\mu\nu}_{\rm eq}$ and write
\begin{eqnarray}
T^{\mu\nu}  = T^{\mu\nu}_{\rm eq} + \delta {T}^{\mu \nu} = T^{\mu\nu}_{\rm eq} + \left(T_{\rm a}^{\mu \nu}- T^{\mu\nu}_{\rm eq} \right). 
\label{TmunuGrada}
\end{eqnarray}
This formula suggests to use the gradient expansion of anisotropic
hydrodynamics in the form
\begin{eqnarray}
\hspace{-0.5cm} T^{\mu\nu} &=& T^{\mu\nu}_{\rm eq} + \hbox{powers of gradients of } T, U^\mu, \xi^{\mu\nu} \hbox{and  } \phi . 
\label{TmunuGrada2}
\end{eqnarray}
Compared to (\ref{TmunuGrad}), the expansion  (\ref{TmunuGrada2}) includes also
the gradients of $\xi^{\mu\nu}$ and $\phi$. On the other hand, similarly to
(\ref{TmunuGrad}),  the expansion  (\ref{TmunuGrada2})  should be done around
the perfect-fluid solution that is determined solely by the conservation
law\rfn{conservation},  see Sec.~\ref{sect:Bjorken}. 

\section{Boost-invariant and transversely homogeneous systems}
\label{sect:binv}

\subsection{Tensor decompositions}

In this work we consider transversely-homogeneous and boost-invariant systems. In this case, following
Refs.~\cite{Florkowski:2011jg,Martinez:2012tu,Tinti:2013vba}, it is convenient
to introduce a basis of four-vectors that can be used to construct all tensor
structures necessary in our analysis:
\begin{eqnarray}
U^\mu &=& (t/\tau,0,0,z/\tau) \, ,
\label{u} \\
Z^\mu &=& (z/\tau,0,0,t/\tau) \, ,
\label{z} \\
X^\mu &=& (0,1,0,0) \, ,
\label{x} \\
Y^\mu &=& (0,0,1,0) \, .
\label{y}  
\end{eqnarray} 
Here $t$ and $z$ are time and space coordinates, and
\mbox{$\tau=\sqrt{t^2-z^2}$} is the (longitudinal) proper time. The time-like
four-vector $U^\mu$ describes the longitudinal Bjorken flow of
matter~\cite{Bjorken:1982qr}. The four-vectors $X$, $Y$, and $Z$ are
space-like and orthogonal to $U$.

We find that 
\begin{equation}
\Delta^{\mu\nu} = g^{\mu\nu} - U^\mu U^\nu = -X^\mu X^\nu - Y^\mu Y^\nu -Z^\mu Z^\nu
\label{Delta}
\end{equation}
and
\begin{equation}
\pi^{\mu\nu} = \pi_\perp \left( X^\mu X^\nu + Y^\mu Y^\nu  \right) +\pi_\parallel  Z^\mu Z^\nu,
\label{pidec}
\end{equation}
where $2 \pi_\perp + \pi_\parallel = 0$ due to the fact that $\pi^{\mu}_{\,\,\,\mu}=0$. This leads to the following decomposition of the energy-momentum tensor for local equilibrium
\begin{eqnarray}
T^{\mu\nu}_{\rm eq} = \varepsilon U^\mu U^\nu + P  \left( X^\mu X^\nu + Y^\mu Y^\nu + Z^\mu Z^\nu \right)
\nonumber \\ \label{Tmunueqdec}
\end{eqnarray}
and for locally anisotropic state
\begin{eqnarray}
T^{\mu\nu} = T^{\mu\nu}_{\rm a} = \varepsilon U^\mu U^\nu + P_\perp \left( X^\mu X^\nu + Y^\mu Y^\nu \right) + P_\parallel Z^\mu Z^\nu.
\nonumber \\ \label{Tmunuadec}
\end{eqnarray}
Here we have introduced the transverse and longitudinal pressure components
\begin{eqnarray}
P_\perp = P + \Pi + \frac{1}{2} \pi_\parallel, \quad P_\parallel = P + \Pi - \pi_\parallel.
\end{eqnarray}

Equations (\ref{TmunuGrada}), (\ref{Tmunueqdec}), and (\ref{Tmunuadec})
indicate that the gradient expansion in our case is defined by the gradient
expansion of the bulk viscosity $\Pi$ and the difference of the longitudinal
and transverse pressure
\begin{eqnarray}
\Delta P = P_\parallel - P_\perp = -\frac{3}{2} \pi_\parallel.
\label{DeltaP}
\end{eqnarray}
For boost-invariant systems, the gradients of scalar quantities such as
$P_\parallel$ or $P_\perp$ are expressed by the derivatives with respect to
the proper time $\tau$. Hence, our task is twofold: we need to find the
dependence of the fluid variables $T$, $\xi^{\mu\nu}$ and $\phi$ on the proper
time $\tau$ and, then, to express the corrections to the energy-momentum
tensor, $\Pi$ and $\Delta P$, in terms of $T$, $\xi^{\mu\nu}$ and $\phi$. In
the conformal limit it is enough to study $\Delta P$, since $\Pi=\phi=0$ in
this case. In Sec.~\ref{sect:anisodist} we argue that our symmetry constraints
allow us to use only one independent scalar anisotropy parameter $\xi$ instead
of the full five-component tensor $\xi^{\mu\nu}$.

\subsection{Bjorken perfect-fluid model}
\label{sect:Bjorken}

In the case where the dissipative terms are neglected we deal with
Eqs.~(\ref{conservation}) only. For boost-invariant and transversally
homogeneous systems, with the flow vector given by Eq.~(\ref{u}), they 
reduce to a single equation of the form~\cite{Bjorken:1982qr}
\beq
\frac{d\eps}{d\tau} = - \frac{\eps+P}{\tau}.
\label{Bjorken1}
\eeq
For conformal systems studied here $\eps = 3 P$, hence, Eq.~(\ref{Bjorken1})
has a solution
\beq
\eps = \eps_0 \left(\frac{\tau_0}{\tau}\right)^{4/3},
\label{Bjorken2}
\eeq
where $\eps_0$ is the energy density at the initial proper time  $\tau_0$, $\eps_0 = \eps(\tau_0)$. 

Since $\eps$ and $P$ describe the system in local equilibrium, we may use the
thermodynamic relations $d\eps = T ds$ and $dP = s \,dT$, where $s$ is the
entropy density. This leads to a scaling solution for the entropy density
\beq
s =  \frac{s_0 \tau_0}{\tau}
\label{BjorkenS}
\eeq
while for the temperature one obtains
\beq
T = T_0 \left(\frac{\tau_0}{\tau}\right)^{1/3}.
\label{BjorkenT}
\eeq
Here $s_0$ and $T_0$ are the initial values of the entropy density and
temperature, respectively. Dissipative effects introduce corrections 
to the above equations
which will be analysed below with the help of the gradient
expansion. 

\section{Boltzmann kinetic equation in the relaxation time approximation}
\label{sect:rta}

\subsection{Kinetic equation}

The kinetic equation in the relaxation time approximation has the form~\cite{Bhatnagar:1954zz}
\beq
p^\mu \p_\mu f(x,p) =  p \cdot U \, \frac{\feq(x,p)-f(x,p)}{\teq},
\label{kineq}
\eeq
where $f(x,p)$ is the one-particle phase-space distribution function depending on particle space-time coordinates $x$ and momenta $p$, and $\feq(x,p)$ is the background equilibrium distribution function. In this work we neglect the effects coming from quantum statistics and set particle masses equal to zero. Thus, in local equilibrium we deal with the Boltzmann distribution function $\feq$ that may be rewritten with the help of the four-vectors (\ref{z})--(\ref{y}) as
\begin{equation}
f_{\rm eq} =  \exp\left(- \frac{1}{T}\sqrt{(p\cdot X)^2 + (p\cdot Y)^2 + (p\cdot Z)^2}\,\, \right),
\label{BoltzmannXYZ}
\end{equation}
where $T$ is the system's (effective) temperature which is determined by the condition that the energy densities obtained with the distributions $f$ and $\feq$ are equal, see \rf{LM0}.

The quantity $\teq$ in (\ref{kineq}) is the relaxation time. In this work we
consider conformal systems where $\teq$ depends on the inverse of the
effective temperature, namely, we use the formula
\beq
\teq = \frac{c}{T},
\label{c}
\eeq
where $c$ is a numerical constant. The calculation of the shear viscosity
coefficient $\eta$ done with Eq.~(\ref{kineq}) leads to the
result~\cite{Anderson:1974q,Czyz:1986mr,Romatschke:2011qp}
\beq
\teq = \frac{5 {\bar \eta}}{T},
\label{crta}
\eeq
where ${\bar \eta}$ is the ratio of the shear viscosity to the entropy
density, ${\bar \eta} = \eta/s$. This means that the coefficient $c$ may be
related directly to the shear viscosity of the system.

We note that for boost invariant and transversally homogeneous systems,
Eq.~(\ref{kineq}) can be solved
exactly~\cite{Baym:1984np,Baym:1985tna,Florkowski:2013lza,Florkowski:2013lya}
and its solutions can be used to assess the agreement between the
kinetic-theory approach and hydrodynamic approaches that have been constructed
with its help~\cite{Florkowski:2013lza,Florkowski:2013lya}. Our present study,
based on the gradient expansion, is complementary to this type of study.

\subsection{Anisotropic distribution functions}
\label{sect:anisodist}


Within the original formulation of anisotropic hydrodynamics~\cite{Florkowski:2010cf,Martinez:2010sc}, one assumes that the non-equilibrium distribution function $f$ is well approximated by the spheroidal Romatschke-Strickland form~\cite{Romatschke:2003ms}, 
\begin{eqnarray}
f _{\rm a}&=&  \exp\left(- \frac{1}{\Lambda}\sqrt{(p\cdot U)^2 + \xi (p\cdot Z)^2}\, \right) \label{RS1} \\
&=&  \exp\left(- \frac{1}{\Lambda}\sqrt{(p\cdot X)^2 + (p\cdot Y)^2 + (1+\xi) (p\cdot Z)^2}\, \right).
\nonumber
\end{eqnarray}
The parameter $\Lambda$ in (\ref{RS1}) defines a typical transverse-momentum scale in the system, while $\xi$ is a scalar anisotropy parameter.  In local equilibrium $\xi \to 0$ and $\Lambda$ can be identified with the temperature $T$.

In Ref.~\cite{Tinti:2013vba} a generalized ellipsoidal parameterization of the anisotropic distribution function was proposed as a good approximation for $f$, namely
\begin{eqnarray}
 f_{\rm a} =   \exp\left(- \frac{E}{\lambda}\, \right),
\label{TF1}
\end{eqnarray}
where
\beq
E^2 = (1+\xi_X) (p\cdot X)^2 + (1+\xi_Y) (p\cdot Y)^2 + (1+\xi_Z) (p\cdot Z)^2. \nn \\
\eeq
The form (\ref{TF1}) becomes important in the case where radial expansion of the system is present \cite{Tinti:2013vba}, since in such cases the pressure anisotropies in the $X$ and $Y$ directions are generally different, which is not included in the original formulation of \AH based on  (\ref{RS1}).  

In this work we consider a boost-invariant and transversely-homogeneous system in which case the two formulations (\ref{RS1}) and (\ref{TF1}) are completely equivalent.  We present (\ref{TF1}) for completeness as several results presented below are obtained within the framework defined in \cite{Tinti:2013vba}.  We note that the anisotropy parameters $\xi_I$ in (\ref{TF1}) satisfy the condition \cite{Tinti:2013vba}
\begin{eqnarray}
\sum_I \xi_I = \xi_X + \xi_Y + \xi_Z = 0 \, .
\label{sumofxis}
\end{eqnarray}
Consequently, the parameterizations (\ref{RS1}) and (\ref{TF1}) are connected through the following set of simple transformations
\begin{eqnarray}
\xi_X &=& \xi_Y = \xi_\perp = -\frac{\xi/3}{1+\xi/3}, \nonumber \\
\xi_Z &=& \xi_\parallel = \frac{2\,\xi/3}{1+\xi/3}, \nonumber \\
\lambda &=& \Lambda (1+\xi/3)^{-1/2}.
\label{RS-TF}
\end{eqnarray}


The form of the anisotropic distribution can be made even more general if we use the expression
\begin{eqnarray}
 f_{\rm a} =   \exp\left(- \frac{1}{\lambda}\sqrt{p_\mu (U^\mu U^\nu+\xi^{\mu \nu} ) p_\nu }\, \right),
\label{AP1}
\end{eqnarray}
where the anisotropy tensor $\xi^{\mu \nu}$ appears. The five parameters in
$\xi^{\mu \nu}$ together with three independent parameters defining the flow
vector $U^\mu$ and $\lambda$ should be taken in such a way that one reproduces
nine independent components of the conformal energy-momentum tensor, see
Eq.~(\ref{TmunuAP}) that represents the anisotropic matching principle
introduced by Tinti in \cite{Tinti:2015xwa}. The latter is a generalisation of
the Landau matching condition which demands that the parameters of the
distribution function reproduce only the energy density and energy flow
vector, see Eq.~(\ref{LandFrameA}).  For a boost-invariant and transversally
homogeneous system, the distribution~(\ref{AP1}) agrees with (\ref{TF1}),
since in this case we have
\begin{eqnarray}
\xi^{\mu \nu} = \xi_\perp \left( X^\mu X^\nu +  Y^\mu Y^\nu \right) + \xi_\parallel  Z^\mu Z^\nu.
\label{ximunu} 
\end{eqnarray}

\section{Anisotropic hydrodynamics equations }
\label{sect:aniso2}

Our last considerations show that with our symmetry constraints we may use the
original Romatschke-Strickland form of the distribution function with a single
anisotropy parameter. We have introduced, however, the forms (\ref{TF1}) and
(\ref{AP1}) because they serve as the starting points for two different
formulations of anisotropic hydrodynamics --- the first one
\cite{Tinti:2013vba} uses the moments of the Boltzmann equation, while the
second one \cite{Tinti:2015xwa} uses the anisotropic matching principle
combined with the exact treatment of the dynamical equations for the pressure
corrections. The latter introduces an infinite set of the coupled equations
that is finally truncated by the assumption that $f = f_{\rm a}$.

In the next two sections we present the form of anisotropic hydrodynamics equations derived in \cite{Tinti:2013vba} and \cite{Tinti:2015xwa}, respectively. Our discussion above suggests that in the two considered cases the hydrodynamic equations can be written as two coupled ordinary differential equations for the functions $T(\tau)$ and $\xi(\tau)$.

\subsection{First option}
\label{sect:ahydro1}

For purely longitudinal and boost invariant expansion, the equations of anisotropic hydrodynamics derived in Ref.~\cite{Tinti:2013vba} can be cast into the following form
\beq
T^4 = \calR(\xi) \Lambda^4,
\label{LM}
\eeq
\beq
\frac{4 \calR(\xi)}{\Lambda} \frac{d\Lambda}{d\tau} + \calR^\prime(\xi) \frac{d\xi}{d\tau} = -\frac{1}{\tau}
\left(\calR(\xi) + \frac{1}{3} \calRL(\xi) \right)
\label{enecon}
\eeq
and
\beq
-\frac{1}{1+\xi} \frac{d\xi}{d\tau} + \frac{2}{\tau} = \frac{\xi}{\teq} \left(\frac{T}{\Lambda} \right)^{5} (1+\xi)^{1/2}.
\label{secondmom}
\eeq
These are three equations for three functions of the proper time: the effective temperature, $T(\tau)$, the transverse-momentum scale, $\Lambda(\tau)$, and the anisotropy parameter, $\xi(\tau)$.

Equations (\ref{LM}) and (\ref{enecon}) follow from the first moment of the kinetic equation\rfn{kineq}, i.e., from the energy-momentum conservation. Equation (\ref{LM}) expresses the condition that the energy density obtained from the anisotropic distribution\rfn{RS1} characterised by $\Lambda$ and $\xi$ is equal to the energy density obtained from the reference equilibrium distribution\rfn{BoltzmannXYZ} characterised by the temperature $T$. The functions $\calR$ and $\calRL$ are defined by the formulas~\cite{Martinez:2010sc}
\beq
\calR(\xi) = \frac{1}{2} \left( 
\frac{1}{1+\xi} + \frac{\hbox{tanh}^{-1}(\sqrt{\xi})}{\sqrt{\xi}}
\right)
\eeq
and
\beq
\calRL(\xi) = \frac{3}{\xi} \left(\calR(\xi) -  \frac{1}{1+\xi} \right).
\eeq

Equation~(\ref{secondmom}) is obtained from the second moment of the kinetic
equation\rfn{BoltzmannXYZ}. Its form is determined by the condition that it
agrees with the \MIS theory for systems that are close to
equilibrium, for more details see derivation of Eqs.~(43), (48) and (57) 
presented in \cite{Florkowski:2014bba}.

Using Eqs.~ (\ref{c}) and (\ref{LM}) in Eqs.~(\ref{enecon}) and (\ref{secondmom}) we obtain two coupled ordinary differential equations for $T$ and $\xi$ only,
\beq
4\,\frac{ \calR(\xi)}{T} \frac{dT}{d\tau} = -\frac{1}{\tau}
\left(\calR(\xi) + \frac{\calRL(\xi)}{3}  \right)
\label{enecon1}
\eeq
and
\beq
-\frac{d\xi}{d\tau} + \frac{2 (1+\xi)}{\tau} = \frac{\xi\,T\,\calR(\xi)^{5/4}}{c}  (1+\xi)^{3/2}.
\label{secondmom1}
\eeq

\subsection{Second option}
\label{sect:ahydro2}

As the second option for the anisotropic hydrodynamics equations we choose the form derived recently in \cite{Tinti:2015xwa}. This form follows from the anisotropic matching principle. One can check that this matching is consistent with Eq. (\ref{enecon1}). On the other hand, Eq.~(\ref{secondmom1}) should be replaced by Eq. (82) from~\cite{Tinti:2015xwa}. In the conformal limit, the latter has the form
\begin{eqnarray}
\frac{d\,\Delta P}{d\tau} = - \frac{T\,\Delta P }{c } - \frac{F}{\tau},
\label{DeltaPeq}
\end{eqnarray}
where $\Delta P$ is the difference of the longitudinal and transverse pressures, see Eq.~(\ref{DeltaP}). Using definitions given in~\cite{Tinti:2015xwa} one finds that  $\Delta P$  can be expressed as
\begin{eqnarray}
\Delta P = -\frac{6 k \pi   \Lambda^4}{\xi} 
 \left(\frac{\xi+3}{\xi+1} + \frac{(\xi-3) \tan^{-1}(\sqrt{\xi})}{\sqrt{\xi}} \right).
 \nonumber \\ \label{DeltaPofy}
\end{eqnarray}
Similarly, one finds the form of the function $F$ appearing on the right-hand
side of (\ref{DeltaPeq}), namely
\begin{eqnarray}
F = -2 (1+\xi) \frac{\partial \Delta}{\partial \xi}.
\nonumber \\ \label{F}
\end{eqnarray}
Using Eq.~(\ref{LM}) to express $\Lambda$ in terms of $T$ and $\xi$ in
(\ref{DeltaP}) and (\ref{F}), and substituting (\ref{DeltaP}) and (\ref{F})
into (\ref{DeltaPeq}) we find an alternative for (\ref{secondmom1}). 

To summarise the last two Sections, we state that we have two options for
anisotropic hydrodynamics equations, these are either Eqs.~(\ref{enecon1}) and
(\ref{secondmom1}), denoted below as \AH I, or Eqs.~(\ref{enecon1}) and
(\ref{DeltaPeq}), denoted below as \AH II. In both cases these are two
ordinary coupled differential equations for the two functions of the proper
time, $T(\tau)$ and $\xi(\tau)$.

\section{M{\"u}ller-Israel-Stewart viscous hydrodynamics}
\label{sect:mis}

The hydrodynamic equations introduced in the previous subsections were
tailored to a specific microscopic theory: kinetic theory in the relaxation
time approximation. In particular, they did not introduce any free parameters
beyond the single parameter, $c$, present already in the microscopic model. In
this Section we briefly review the complementary approach, which assumes
essentially nothing beyond thermodynamic equilibrium in the far future. In
particular, no quasiparticle picture or specific microscopic model is adopted,
yielding a very powerful, general theory. This generality, as we shall see,
comes at a price: we cannot expect this theory to provide as fine a picture 
as that provided by hydrodynamic equations developed to mimic a specific
microscopic model.

As described in \rfs{sect:hydro}, to formulate a theory of dissipative
relativistic hydrodynamics in the conformal cases one needs to provide five
equations in addition to the four conservation equations\rfn{conservation}.
The \IS theory postulates that these should take the form of relaxation
equations~\cite{Muller:1967zza,Israel:1976tn,Israel:1979wp}
\bel{relax}
\Delta^{\alpha\beta}_{\mu\nu} U^\gamma \p_\gamma \pi^{\mu \nu} 
= - \f{1}{\tau_\pi} \left(\pi^{\alpha\beta} - 2 \eta
\sigma^{\alpha\beta} \right) + \ldots  
\ee
where $\tau_\pi$ is the relaxation time. This guarantees that the shear stress
tensor approaches the Navier-Stokes form \rfn{NavierStokes} at late times. 
It is easy to see that \rf{relax} can be solved iteratively obtaining 
\bel{misgrad}
\pi^{\alpha\beta}  =  2 \eta \sigma^{\alpha\beta} + \ldots \,,
\ee
where the ellipsis contains terms of second and higher order in gradients. It is
clear that all orders in gradients will appear here. 

The formulation of \MIS theory in \rfc{Baier:2006um} rests on the observation
that by a judicious choice of terms on  
the RHS of \rf{relax} one can generate all possible terms up to second order
in gradients with coefficients which can be chosen at will. In this way, by a
suitable choice of these coefficients one can match a calculation of the
energy-momentum tensor expectation value in any microscopic theory or model up
to second order in gradients. Of course, all higher orders will typically not
be matched, but sufficiently close to equilibrium this is not an issue. It is
instructive to recall at 
this point that the formulation of the \MIS equations was
prompted by a calculation of the gradient expansion of the energy-momentum
tensor performed in ${\cal N}=4$ supersymmetric Yang-Mills theory
\cite{Bhattacharyya:2008jc} using the AdS/CFT correspondence. It turned out
that the original \IS theory did not have all the terms necessary to match the
result of the microscopic calculation.

The complete set of \MIS equations for conformal hydrodynamics given in
\rfc{Baier:2006um} involves 5 second order transport
coefficients $\tau_\Pi, \lambda_1, \lambda_2, \lambda_3,\kappa$. 
If boost invariance is imposed, the resulting \MIS equations involve only
$\tau_\Pi, \lambda_1$, and the shear viscosity:
\beal{miseqn}
\tau  \dot{\epsilon} &=& - \frac{4}{3}\epsilon + \phi\nonumber\, , \\
\tau_\Pi \dot{\phi} &=& 
\frac{4 \eta}{3 \tau } 
- \frac{\lambda_1\phi^2}{2 \eta^2}
- \frac{4 \tau_\Pi\phi}{3 \tau }
- \phi \, ,
\eea
where the dot denotes a proper time derivative and $\phi\equiv-\pi_\parallel$
is the single independent component of the shear stress tensor.

For comparison, we also analyse below: i)  the hydrodynamic equations derived
in \rfc{Denicol:2012cn},  which for the case of the Bjorken flow differ from
(\ref{miseqn}) by the second line, which reads
\beal{dnmreqn}
\tau_\Pi \dot{\phi} &=& 
\frac{4 \eta}{3 \tau } 
- \frac{38}{21} \f{\tau_\Pi\phi}{\tau }
- \phi \, ,
\eea
and ii) a truncated version of the second line in (\ref{miseqn}) 
\beal{iseqn}
\tau_\Pi \dot{\phi} &=& 
\frac{4 \eta}{3 \tau } 
- \frac{4 \tau_\Pi\phi}{3 \tau }
- \phi \, ,
\eea
which can be connected with one of the early versions of the Israel-Stewart theory.
The results obtained with (\ref{dnmreqn}) and (\ref{iseqn}) will be denoted by the labels 
\DNMR and \IS\!\!, respectively.  In the following Section, it will be clarified in what sense 
Eqs.~(\ref{miseqn}), (\ref{dnmreqn}), and (\ref{iseqn}) are consistent with each other.

\section{Dimensionless gradient expansion}
\label{sect:gradaniso}

In this Section we study the gradient expansion for the Bjorken flow, which
amounts to calculating gradient corrections to the perfect-fluid Bjorken
solution~\cite{Bjorken:1982qr}. Thus, it is convenient, as seen in a number of
recent studies, to consider the dimensionless function $g$ of a
dimensionless variable $w$, namely\footnote{In \cite{Heller:2015dha} this
function was called $f$.}
\bel{gow}
g = \f{1}{T}\f{d w}{d\tau}, \qquad w = \tau T.
\ee
This quantity is related in a trivial way to the dimensionless pressure
anisotropy
\be
\Delta = \frac{\Delta P}{P} = 3 \frac{P_\parallel-P_\perp}{\varepsilon} =  12 \left(g - \f{2}{3}\right).
\ee
The gradient expansion for boost-invariant flow takes the form of an expansion
\bel{hydroser}
g(w) = \sum_{n=0}^\infty g_n w^{-n},
\ee
where $g_0=2/3$, which corresponds to the perfect-fluid behavior
$T\sim\tau^{-1/3}$~\cite{Bjorken:1982qr}. The fact that only integer powers
appear in \rf{hydroser} is a very convenient feature. 

In Ref.~\cite{Heller:2015dha}, where the \MIS theory was considered, it was
possible to find a closed first order ordinary differential equation satisfied
by the function $g(w)$. The gradient expansion was then calculated by looking
for a solution in the form of a series in $1/w$. In the present case, it
appears difficult to find a closed equation for $g(w)$, so we 
first calculate the expansions of $T$ and $\xi$ in powers of $\tau$ and then
determine the coefficients $g_n$ by solving \rf{gow} as an expansion of the
form~(\ref{hydroser}).

It is convenient to write the series for $T(\tau)$ and $\xi(\tau)$ in the
forms
\bel{Tgradser}
T = T_0 \left(\frac{\t0}{\tau}\right)^{1/3} \left( 1 + \sum_{n=1}^\infty
\left(\f{c}{T_0\tau_0}\right)^n \tn \left(\frac{\t0}{\tau}\right)^{2 n/3} \right),
\ee
\bel{xigradser}
\xi(\tau) = \sum_{n=1}^\infty \left(\f{2 c}{\tau_0 T_0}\right)^n \xi_n
\left(\f{\tau_0}{\tau}\right)^{2n/3} ,
\ee
where $T_0$ is the initial temperature at some initial proper time $\t0$ for
the Bjorken solution~\cite{Bjorken:1982qr}. We insert (\ref{Tgradser}) and
(\ref{xigradser}) into either Eqs.~(\ref{enecon1}) and (\ref{secondmom1}) or
Eqs.~(\ref{enecon1}) and (\ref{DeltaPeq}).  These two pairs of equations
correspond to two different options of constructing anisotropic hydrodynamics,
see Sections~(\ref{sect:ahydro1}) and (\ref{sect:ahydro2}), respectively.  The
form of these equations is cumbersome, but they are first order differential
equations which can be solved order by order in powers of $\tau^{2/3}$ to
determine the values of the coefficients $\xi_n$ and $t_n$.

For the first option of anisotropic hydrodynamics based on Eqs.~(\ref{enecon1}) and (\ref{secondmom1}), one finds 
\bel{xin1}
(\xi_n) = \left(1, \frac{11}{15}, \frac{622}{1575}, \frac{893}{10125}, \dots\right) , \qquad n = 1,2,\dots
\ee
and
\bel{tn1}
(t_n) = \left(1, -\frac{2}{15}, -\frac{2}{315}, \frac{268}{14175}, \dots\right), \qquad n = 0,1,\dots ,
\ee
while for the second option based on Eqs.~(\ref{enecon1}) and (\ref{DeltaPeq}) we get
\bel{xin2}
(\xi_n) = \left(1, \frac{82}{105}, \frac{782}{1575}, \frac{8325224}{38201625}, \dots\right) , \qquad n = 1,2,\dots
\ee
and
\bel{tn2}
(t_n) = \left(1, -\frac{2}{15}, -\frac{4}{315}, \frac{1208}{99225},\dots\right), \qquad n = 0,1,\dots
\ee

The results (\ref{xin1}) and (\ref{tn1})  imply the following coefficients in \rf{hydroser}
\bel{gn1}
(g_n) = \left( \frac{2}{3}, \frac{4c}{45} , \frac{8c^2}{945} , -\frac{184c^3}{4725}\dots\right) , \qquad
n = 0,1,\dots
\ee
Similarly, the series (\ref{xin2}) and (\ref{tn2})  imply
\bel{gn2}
(g_n) = \left(\frac{2}{3}, \frac{4c}{45}, \frac{16 c^2}{945}, -\frac{176c^3}{6615}, \dots\right) , \qquad
n = 0,1,\dots
\ee
The leading term is the correct perfect fluid value. The first subleading
term represents the viscous correction. The most general form of the energy
momentum tensor in conformal hydrodynamics \cite{Baier:2007ix} implies that 
\be
\f{\eta}{s} = \f{9}{4} g_1 
\ee
so in the theory considered here $\eta/s = c/5$. This value is already known
from other
considerations~\cite{Anderson:1974q,Czyz:1986mr,Romatschke:2011qp}. It is also
confirmed by comparing our results with the results of a direct computation of
the gradient expansion of kinetic theory in the~\RTA \cite{hks}.

It is interesting to compare the gradient expansion of \AH\, to the underlying
kinetic theory, as well as to the \MIS theory. The leading terms of the
gradient expansions are listed in Table~I (we have set the coefficient $c=1$
for readability; it can be restored using the fact that $g_k\sim c^k$).
For Bjorken flow \MIS has effectively three parameters ($\tau_\Pi,
\eta,\lambda_1$) so one can adjust them to reproduce the exact computation in
the RTA model up  to second order.

The only free parameter of \AH for conformal systems is a constant $c$ fixing
the viscosity to entropy density ratio. It appears in the same way in \AH as
in the underlying kinetic equation through Eq.~(\ref{c}) that fixes the
relaxation time. The agreement of the first order terms in the gradient
expansion for \AH and the kinetic theory shows that \AH (in the two considered
versions) properly includes the effects of shear viscosity.

In the second order, the first version of \AH misses the \RTA results by a
factor of two yielding $8c^2/945$ instead of $16c^2/945$. Very interestingly,
the second version of \AH reproduces exactly the \RTA result. This gives
further support for anisotropic hydrodynamics based on the anisotropic
matching principle~\cite{Tinti:2015xwa}.

At third order however both MIS and \AH depart from the kinetic theory
results, but \AH is significantly closer. Numerically \MIS gives 
39\% of the \RTA result, \AH I gives 88\%, while \AH II gives 60\%.

\begin{center}
\begin{table}[t]
  \label{tab:table1}
  \begin{tabular}{|c|c|c|c|c|}
    \hline
    $n$ & \RTA & \MIS & \AH I & \AH II \\
    \hline
    $0$ & $2/3$ & $2/3$ & $2/3$ & $2/3$ \\
    \hline
    $1$ &  $4/45$ &  $4/45$ & $4/45$ & $4/45$ \\
    \hline    
    $2$ & $16/945 $ & $16/945$ & $8/945$ & $16/945 $\\
    \hline
    $3$ & $- 208/4725$ & $-1712/99225$  & $-184/4725$ & $-176/6615$\\
    \hline
  \end{tabular}
    \caption{Leading coefficients of gradient expansions for \RTA\!\!,
      \MIS\!\!, \AH I  and \AH II.}
\end{table}
\end{center}

\begin{center}
\begin{table}[t]
  \label{tab:table2}
  \begin{tabular}{|c|c|c|c|c|}
    \hline
    $n$ & \RTA & \MIS & \DNMR & \IS \\
    \hline
    $0$ & $2/3$ & $2/3$ & $2/3$ & $2/3$ \\
    \hline
    $1$ &  $4/45$ &  $4/45$ & $4/45$ & $4/45$ \\
    \hline    
    $2$ & $16/945 $ & $16/945$ & $16/945$ & $8/135 $\\
    \hline
    $3$ & $- 208/4725$ & $-1712/99225$  & -$304/33075$ & $112/2025$\\
    \hline
  \end{tabular}
    \caption{Leading coefficients of gradient expansions for \RTA\!\!,
      \MIS\!\!, \DNMR  and \IS\!\!.}
\end{table}
\end{center}

In Table~II we show the leading coefficients of gradient expansions for the
kinetic theory model and various hydrodynamic approaches (\RTA\!\!, \MIS\!\!,
\DNMR and \IS\!\!, respectively). It is interesting to observe that \MIS and
\DNMR agree up to the second order. This is expected, since both \MIS and
\DNMR have been constructed as consistent expansions. They both agree with
\RTA\!\!: the first because the parameters $\tau_\Pi, \eta,\lambda_1$ of \MIS
have been adjusted to reproduce the \RTA result, the second because the
kinetic coefficients used in (\ref{dnmreqn}) have been obtained directly from
the \RTA kinetic equation~\cite{Denicol:2012cn}.

The \IS results differ from those obtained for \RTA\!\!, \MIS and \DNMR already at
the second order. Note that the old \IS theory has $g_3$ of
the wrong sign. This points to the importance of a nonzero value of the 
$\lambda_1$ term in (\ref{miseqn}).  The poorer performance of \IS compared to
\DNMR in the gradient expansion is similar to the situation described in
\cite{Florkowski:2013lya}.  In this work solutions of (\ref{iseqn}) and
(\ref{dnmreqn}) were compared with the exact solutions of the kinetic theory
indicating that the \DNMR approach better reproduces the kinetic-theory
results.

\section{Large order behavior}
\label{sect:Borel}

It has recently become clear that large order behavior of gradient
expansions contains important information about the non-hydrodynamic sector of
the theory. This is the case both at the level of microscopic theories
\cite{Heller:2013fn,hks} and at the level of hydrodynamics
\cite{Heller:2015dha,Aniceto:2015mto}. Since hydrodynamics can be treated as
an effective description of microscopic systems, one may aim not only to match
the low orders of the gradient expansion, but also the large order behavior,
which is tantamount to matching the non-hydrodynamic sectors. Of course, one
may choose to refrain from using the hydrodynamic description when dependence
on the non-hydrodynamic sector (the ``regulator sector'' in the language of
\rfc{Spalinski:2016fnj}) is nontrivial, in which case only matching of the low
orders would be important. In this section, however, we will examine the large
order behavior of the formal gradient expansions of anisotropic hydrodynamics
to determine what kind of non-hydro modes these theories contain.

Using the methods described in the previous section one may, with relatively
modest effort, determine the coefficients $g_n$ to order 140 in both models of
anisotropic hydrodynamics considered in this paper.  Examination of
these coefficients shows that the series has vanishing radius of convergence,
with $g_n\sim n!$. For the series~\rfn{gn2}, this is illustrated in
Fig.~\ref{fig:div} below; the result for the series~\rfn{gn1} is
analogous. 

\begin{figure}[t]
\center
\includegraphics[height=0.25\textheight]{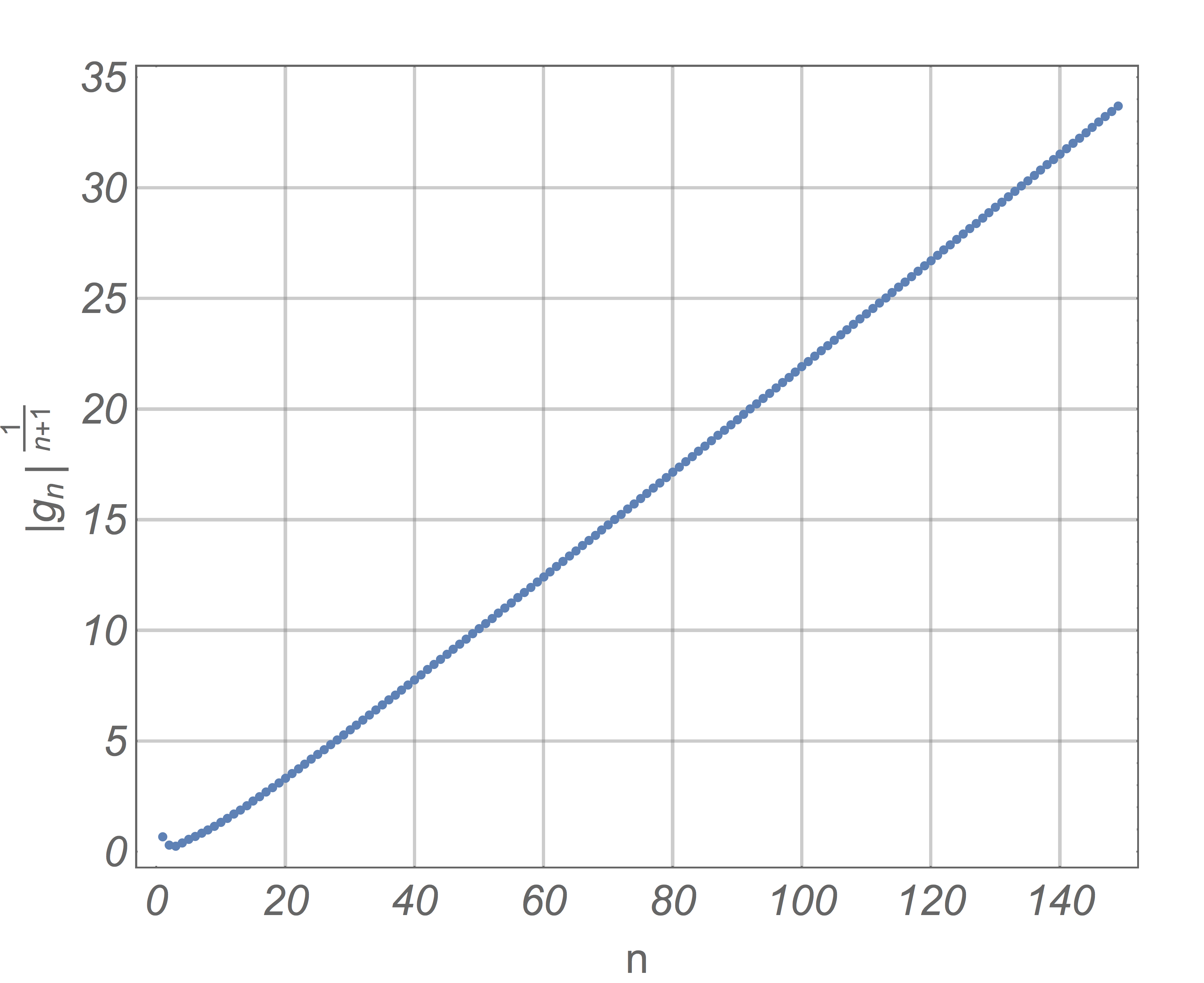}
\caption{The coefficients in \rf{gn2} grow as $n!$ }
\label{fig:div}
\end{figure}

\begin{figure}[t]
\center
\includegraphics[height=0.25\textheight]{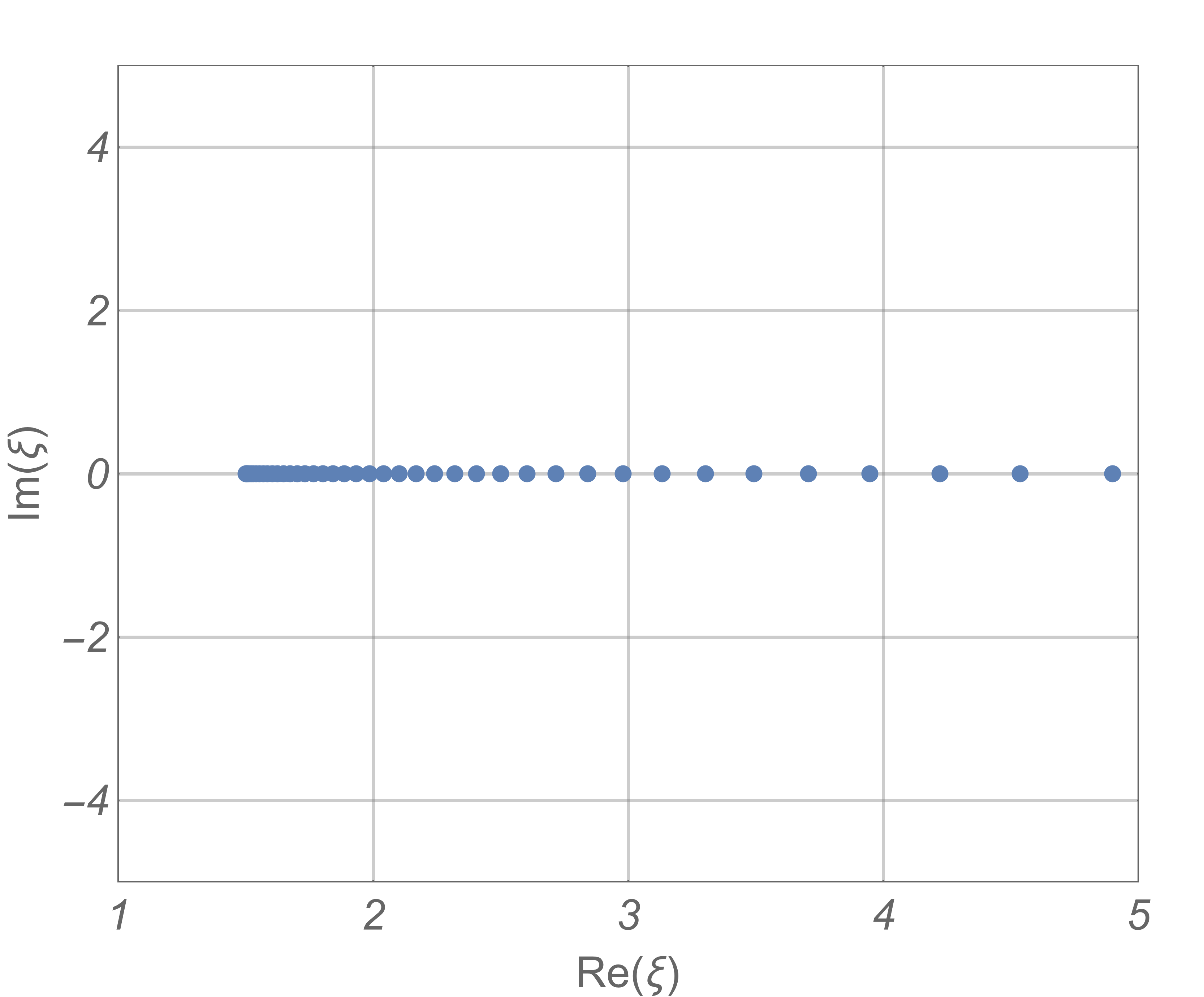}
\caption{Poles of the symmetric Pade approximant to the Borel transform of the
  series $\{g_n\}$ in \rf{gn2}. }
\label{fig:poles}
\end{figure} 

As in \cite{Heller:2015dha,Basar:2015ava,Aniceto:2015mto}, we will study the
singularities of the analytic continuation of the Borel transform as a means
to learn about non-hydrodynamic modes of the theory. The Borel transform of $g$
is given by
\bel{borel}
g_B(\xi) = \sum_{n=0}^\infty \frac{g_n}{n!} \xi^n,
\ee
and represents a series which has a finite radius of convergence.  The
analytic continuation of series \rfn{borel}, denoted by $\tilde{g}_B(\xi)$,
would be needed to invert the Borel transform via the formula
\be
\label{borelsum}
g_{R}(w) =  w \int_{C} d\xi \, e^{- w \xi} \, \tilde{g}_B(\xi),
\ee
where $C$ denotes a contour in the complex plane connecting $0$ and $\infty$.
We perform the analytic continuation using diagonal Pad\'e approximants of
order 70. This function has a sequence of poles along the positive real axis,
starting at $\xi_0=1.500$, which signals the presence of a cut originating at
that point. A consequence of this is a ``nonperturbative'' ambiguity of the
same kind as that seen in \cite{Heller:2015dha}. The implication of this is
that the hydrodynamic series must be regarded as the lowest order element of a
transseries. This line of reasoning can be continued as in
\rfc{Heller:2015dha} (see also \cite{Basar:2015ava,Aniceto:2015mto}). The main
conclusion for our present purposes is however that the cut along the
real axis indicates the presence of a single 
non-hydrodynamic mode, which is purely decaying, as in the \MIS theory.
The location of the start of the cut determines the rate of exponential
decay to be $3/2 c$.

\section{Summary}
\label{sect:summary}

We have examined the gradient series solutions in two formulations of
anisotropic hydrodynamics and compared them to similar results in \MIS theory,
as well as the gradient series for the model of kinetic theory in the
relaxation time approximation which is the underlying microscopic theory for
the anisotropic hydrodynamics considered here. 

The gradient expansions in hydrodynamic theories are divergent, so their
usefulness, apart from formal comparisons, lies in the fact that keeping only
a few leading terms gives a reasonable approximation at late times. The theory
of asymptotic series provides the concept of optimal truncation, which in the
cases considered is of the order of a few terms. When comparing with a
microscopic model, such as the kinetic theory under consideration here, it is
interesting to ask how many terms in such an expansion should a
hydrodynamic description aim to capture. A conservative point of view would be
to assume that the first two orders are the most relevant, so that one should
determine the coefficients which enter the \MIS equations and use
these. However, experience from the numerical study of
\rfc{Florkowski:2013lya} suggests that one can do better, and the evidence
provided by the gradient expansions studied in this paper suggests that
matching higher orders is perhaps indicative of better numerical
performance. One should however keep in mind that a given initial condition
typically involves non-hydrodynamic modes along with hydrodynamic ones, 
and while the decay of the former is exponential, they will affect early time
evolution. Thus conclusions based on low orders of the gradient expansion 
are relevant only for sufficiently late times.

We have also studied the large-order behavior of the gradient series of
anisotropic hydrodynamics, establishing that the series is divergent, as in
the two cases studied previously \cite{Heller:2015dha,Aniceto:2015mto}. The
singularities of the Borel transform indicate that this  theory involves a single, purely
decaying non-hydrodynamic mode, very much like what is found in MIS theory. This
suggests that the pattern of attaining the hydrodynamic attractor should be
similar in both cases.

\bigskip
{\bf Acknowledgments:} We thank Micha\l{} Heller and Michael Strickland for  
discussions concerning the manuscript, and Gabriel Denicol for clarifying comments
about Eq.~(\ref{dnmreqn}). W.F. was supported by Polish National Science Center
Grant DEC-2012/06/A/ST2/00390, M.S. was supported by the Polish National
Science Centre Grant 2015/19/B/ST2/02824, and R.R. was supported by Polish 
National Science Center Grant DEC-2012/07/D/ST2/02125.

\bibliography{transport}

\end{document}